\documentclass{PoS}
\usepackage{cite}
\usepackage{amsmath}
\usepackage{amssymb}

\title{Two-flavor QCD at finite quark or isospin density}

\ShortTitle{Two-flavor QCD at finite quark or isospin density}

\author{P. Cea\\
            Dipartimento di Fisica dell'Universit\`a di Bari, I-70126 Bari, Italy and INFN, Sezione di Bari, I-70126 Bari, Italy\\
            E-mail: \email{paolo.cea@ba.infn.it}}
           
\author{\speaker{L. Cosmai}\\
        INFN, Sezione di Bari, I-70126 Bari, Italy\\
        E-mail: \email{leonardo.cosmai@ba.infn.it}}
            
\author{M. D'Elia\\
            Dipartimento di Fisica dell'Universit\`a di Pisa and INFN - Sezione di Pisa, Largo Pontecorvo 3, 56127 Pisa, Italy\\
            E-mail: \email{delia@df.unipi.it}}

\author{A. Papa\\
            Dipartimento di Fisica dell'Universit\`a della Calabria, I-87036 Rende (Cosenza), Italy and INFN, Gruppo collegato di Cosenza, I-87036 Rende (Cosenza), Italy\\
            E-mail: \email{papa@cs.infn.it}}

\author{F.  Sanfilippo\\
            Laboratoire de Physique Th\'eorique (Bat. 210) Universit\'e Paris
            SUD, F-91405 Orsay-Cedex, France \\
             E-mail: \email{francesco.sanfilippo@th.u-psud.fr}}

\abstract{We exploit analytic continuation to prolongate to the region of real chemical potentials the (pseudo)critical lines of QCD with two degenerate staggered fermions at nonzero temperature and quark or isospin density obtained in the region of imaginary chemical potentials. We determine the curvatures at zero chemical potential and quantify the deviation between the cases of finite quark and of finite isospin chemical potential. In both circumstances deviations from a quadratic dependence of the pseudocritical lines on the chemical potential are clearly seen. We try different extrapolations and, for the nonzero isospin chemical potential, confront them with the results of direct Monte Carlo simulations. We also find that, as for the finite quark  chemical potential, an imaginary isospin chemical potential can strengthen the transition till turning it into strong first order.}

\FullConference{The 30 International Symposium on Lattice Field Theory - Lattice 2012,\\
		June 24-29, 2012\\
		Cairns, Australia}

\begin{document}

\section{Introduction}
The determination of the QCD phase diagram in the temperature-quark density plane is becoming increasingly important, due to its impact in cosmology and in the physics of compact stars and of heavy-ion collisions. 
The first-principle nonperturbative approach of discretizing QCD on a space-time lattice and performing numerical Monte Carlo simulations is plagued, at nonzero quark chemical potential, by the well-known sign problem: the fermionic determinant is complex and the Monte Carlo sampling becomes unfeasible.
Analytic continuation is amongst the possible alternatives to solve (approximately) the sign problem\cite{Alford:1998sd,deForcrand:2002yi,D'Elia:2002gd}.
It consists in performing Monte Carlo simulations at imaginary chemical potential $\mu = i \mu_{\text{IM}}$: where there is no sign problem. 
The results obtained at imaginary chemical potential are then analytically prolongated ($\mu_{\text{IM}} \rightarrow -i \mu$) at real values of the chemical potential.
There are however limitations due to ambiguity in the interpolation and nonanalyticities and periodicity\cite{Roberge:1986mm},
so that reliable estimations are expected only for $\text{Re}(\mu)/T \lesssim 1$, where $T$ is the temperature.

In previous works~\cite{Cea:2006yd,Cea:2007vt,Cea:2009ba} we have studied the analytical continuation of the pseudo critical line in the case of  SU(2) with $n_f=8$ staggered fermions and finite quark density and SU(3) with $n_f=8$ staggered fermions and finite isospin density.
It was found that the nonlinear terms in the dependence of $\beta_c$ on $\mu^2$  in general cannot be neglected and that the extrapolation to real $\mu$ may be wrong otherwise. We have also studied~\cite{Cea:2010md} SU(3) with $n_f=4$ staggered fermions and finite quark density. In this case we observed deviations in the pseudocritical line from the linear behavior in $\mu^2$ for larger absolute values of $\mu^2$ and we saw that there are several possible extrapolations to real $\mu$ that are in agreement with each other up to $\mu/T \simeq 0.6$.

In the present study~\cite{Cea:2012ev} we consider two-flavor QCD in presence of a quark or an isospin  chemical potential in the standard staggered discretization for fermion fields, whose partition function, in the standard staggered discretization for the fermion fields, reads 
\begin{equation}
\label{zeta}
Z_{q/{\rm iso}}(T,\mu) 
\equiv \int \mathcal{D}U e^{-S_{G}} 
(\det M [\mu])^{1\over 4} 
(\det M [\pm \mu])^{1\over 4}  \,.
\end{equation}
In Section 2 we present results on the analytic continuation of the critical line,   $T_c(\mu)$  from imaginary to real  $\mu$  in the case of a finite  isospin chemical potential $\mu_{\text{iso}}$, where simulations are available  for both imaginary and real  $\mu_{\text{iso}}$ and on the analytic continuation of the quark chemical potential $\mu_q$.
In Section 3 we make a comparison between the two theories at finite  $\mu_q$    or  $\mu_{\text{iso}}$, quantifying systematic differences for quantities like the curvature of the pseudocritical line at zero chemical potential. In Section 4 we study the nature of the transition as a function of the isospin chemical potential.

\section{Analytic continuation of the pseudocritical line}
\label{pseudocritical}

\begin{figure}[htbp]
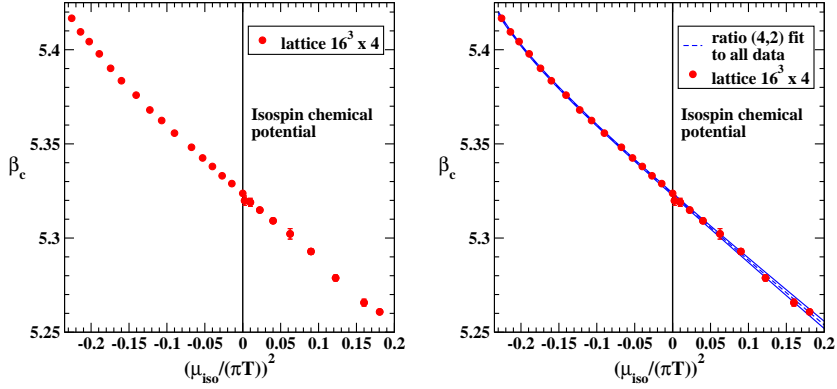
 
\centering
\begin{tabular}{cc}
\includegraphics[width=.35\textwidth,clip]{./FIGURES/peaks_iso_senzafit.eps} &
\includegraphics[width=.35\textwidth,clip]{./FIGURES/peaks_iso.eps} 
\end{tabular} 
\caption{(Left) Pseudocritical couplings in finite isospin SU(3) with $n_f=2$. Negative values of  $(\mu_{\text{iso}}/(\pi T))^2$ correspond to imaginary chemical potentials. (Right) Fit to real and imaginary isospin chemical potential data 
using a ratio of a fourth to second-order polynomial (ratio (4,2) fit).}
\label{fig1}
\end{figure}

\begin{figure}[htbp]
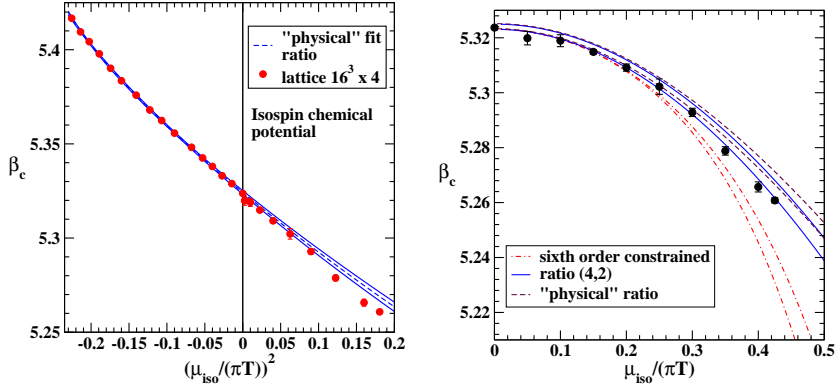
 
\centering
\begin{tabular}{cc}
\includegraphics[width=.35\textwidth,clip]{./FIGURES/SU3_Nf2_iso_physical.eps} &
\includegraphics[width=.35\textwidth,clip]{./FIGURES/SU3_Nf2_iso_extrapolations_lin.eps} 
\end{tabular} 
\caption{(Left) Fit to the pseudo critical couplings in finite isospin SU(3) with $n_f=2$ according to the "physical" fit Eq.~(2.3). 
(Right) Extrapolation to real isospin chemical potentials of the sixth-order constrained, ratio fourth to second-order polynomials and "physical"  fits 
(only the border of the 95\% CL band have been reported). Data points (circles) are the results of Monte Carlo simulations performed directly at real isospin chemical potential.}
\label{fig2}
\end{figure}

We performed numerical simulations on a $16^3\times4$   lattice   (apart from some special cases where we varied the spatial size to investigate the critical behavior) 
for bare quark mass $am=0.05$ corresponding to $m_\pi \sim 400 \, {\text {MeV}}$.
We used the Rational Hybrid Monte Carlo (RHMC) algorithm, properly modified for the inclusion of quark/isospin chemical potential.
Typical statistics have been  around 10k trajectories of 1 Molecular Dynamics unit for each run,  growing up to 100k trajectories for 4--5  $\beta$ values around the pseudocritical point, for each $\mu^2$,  in order to correctly sample the critical behavior at the transition.
The pseudocritical $\beta(\mu^2)$  has been determined as the value for which the susceptibility of the (real part of the) Polyakov loop exhibits a peak. 
In fig.~\ref{fig1}  (left) the data for the pseudocritical coupling versus $(\mu_{\text{iso}}/(\pi T))^2$ are shown. 
To interpolate  these  values  we can exploit the ratio of polynomials:
\begin{equation}
\label{ratio}
\beta_c (\mu^2) = \frac{a_0 + a_1 (\mu/(\pi T))^2 + a_2 (\mu/(\pi T))^4
+ a_3 (\mu/(\pi T))^6}{1 + a_4 (\mu/(\pi T))^2} \,.
\end{equation}
The fit to all data requires (see fig.~\ref{fig1}) at least a ratio of fourth order to second order polynomial (ratio (4,2) fit) and gives a  $\chi^2/{\text{d.o.f.}} = 0.6$.
If we consider data with  $(\mu/\pi T)^2 \ge -0.375^2$  a linear (in $(\mu/\pi T)^2$) polynomial works quite well ($\chi^2/{\text{d.o.f.}} = 0.95$), contrary to our previous findings for other theories\cite{Cea:2006yd,Cea:2007vt,Cea:2009ba,Cea:2010md}  where nonlinear corrections are more important for imaginary values than for real ones.
The interpolation to only imaginary $(\mu_{\text{iso}}/(\pi T))$ data using a atio (4,2) fit  gives a  $\chi^2/{\text{d.o.f.}} = 0.49$. 
We have also interpolated 
imaginary isospin chemical potential data using  the  implicit relation between $\beta_c$  and $\mu^2$ 
\begin{equation}
\label{betacmu2}
 a^2(\beta_c(\mu^2))|_{\rm 2-loop}  = a^2(\beta_c(0))|_{\rm 2-loop}  
  \times \frac{1+A\,x + B\,x^2}{1+C\,x}
\end{equation}
and  the following interpolating function ("physical" fit) given in terms of the physical units $x\equiv\mu/(\pi T)$ and $T/T_c(0)$,
\begin{equation}
\label{Tc0Tcmu}
\left[\frac{T_c(0)}{T_c(\mu)}\right]^2=\frac{1+A\,x +B\,x^2}{1+C\,x} \,,
\end{equation}
with $T=1/(N_t a(\beta))$,  we also get a very good $ \chi^2/{\text{d.o.f.}}=0.53$ and a corresponding  prolongation to real values that works quite well
(see fig.~\ref{fig2} left).
Another quite good interpolation of imaginary isospin data is attained by means of a sixth-order constrained polynomial fit, where the coefficient of  $(\mu/(\pi T))^2$
is fixed at the value derived from a linear (in $(\mu/(\pi T))^2$) fit at small imaginary chemical potential data.
In fig.~\ref{fig2} (right) the extrapolations to real isospin chemical potentials together with results from simulations at real values are shown.
We can see that several  extrapolations agree up to $\mu/(\pi T) \lesssim 0.2$. 
Therefore we may conclude that
different interpolations that well reproduce imaginary data, lead to distinct extrapolations (as we have seen~\cite{Cea:2010md}  for $n_f=4$ SU(3)).

\begin{figure}[htbp]
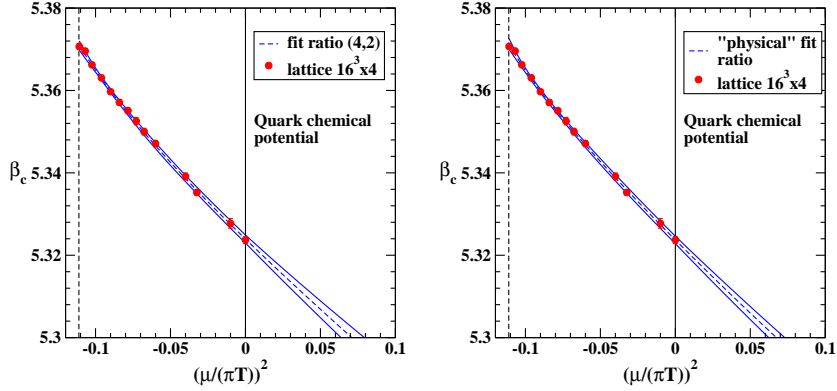
 
\centering
\begin{tabular}{cc}
\includegraphics[width=.35\textwidth,clip]{./FIGURES/SU3_Nf2_quark_ratio42.eps} &
\includegraphics[width=.35\textwidth,clip]{./FIGURES/SU3_Nf2_quark_physical.eps} 
\end{tabular} 
\caption{Fits to the pseudocritical couplings at finite quark density:
ratio of a 4th- to 2nd-order polynomial (left) and ``physical'' fit according 
to the function~(2.2) (right). The dashed vertical line
indicates the boundary of the first Roberge-Weiss sector, $(\mu_{\text{IM}})/(\pi T)=1/3$.}
\label{fig3}
\end{figure}

\begin{figure}[htbp]
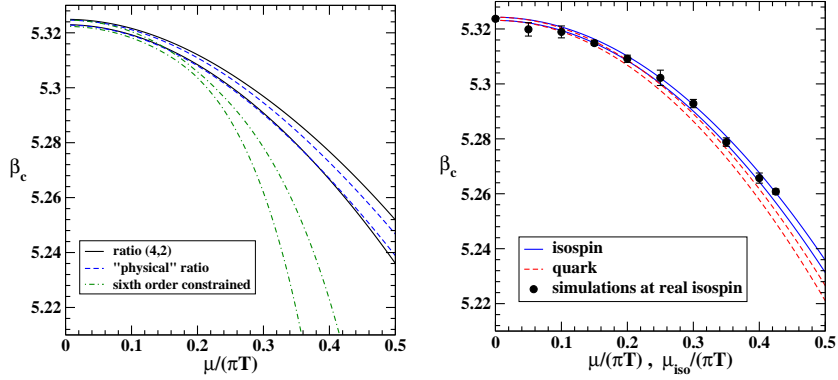
 
\centering
\begin{tabular}{cc}
\includegraphics[width=.35\textwidth,clip]{./FIGURES/SU3_Nf2_quark_extrapolations_lin.eps} &
\includegraphics[width=.35\textwidth,clip]{./FIGURES/SU3_Nf2_iso_linear-small.eps} 
\end{tabular} 
\caption{(Left) Extrapolation to real quark chemical potentials of the 6th-order 
constrained, ratio (4,2) of polynomials and "physical" ratio 
fits (only the borders of the 95\% CL band have been
reported).
(Right) Comparison between the extrapolations to real quark and isospin 
chemical potential of the fits linear in $(\mu/(\pi T))^2$. Data points
(circles) are the results of Monte Carlo simulations performed directly 
at real isospin chemical potential.}
\label{fig4}
\end{figure}

Let us move now to the nonzero quark chemical potential simulations. In this case the sign problem prevents us to perform simulations at real values of the quark chemical potential. In fig.~\ref{fig3} (left) we can  see that  the ratio (4,2) interpolation used in the case of isospin chemical potential is well suited here too,  giving a $\chi^2/{\text{d.o.f.}} = 0.60$. If we tried a linear fit (in $\mu^2$)  we got  a largely unsatisfactory $\chi^2/{\text{d.o.f.}} = 2.87$.
As  for  the isospin chemical potential we also tried  the "physical" fit (Eq.~(\ref{Tc0Tcmu}) to the imaginary quark chemical potential data.
The result, as shown in fig.~\ref{fig3} (right), is  good also in the present case ($\chi^2/{\text{d.o.f.}}=0.51$). 
Assuming that it is possible to extrapolate down to $T=0$ the relation $T_c(\mu)/T_c(0)$ versus $\mu$ (Eq.~(\ref{Tc0Tcmu}),
we get the following estimate for the chemical quark potential at zero temperature:
\begin{equation}
\label{mucritical}
\mu_c(T=0)=\sqrt{\frac{C}{B}} T_c(0) = 3.284(65) T_c(0) \,,
\end{equation}
to be compared with $\mu_c(T=0)= 2.73(58) T_c(0)$   of ref.~\cite{PhysRevD.83.114507} with $n_f=2$ Wilson fermions.
In fig.~\ref{fig4} the extrapolations at real values of the quark chemical potential  starting from three different  successful  interpolating functions at imaginary chemical potential values are compared.  The three  analytic continuations   begin to deviates at $\mu/(\pi T) > 0.1$. 
However  two of these extrapolations (in particular the ratio of polynomials and the "physical" fit) continue to be in good agreement. Moreover we 
observe that in the case of isospin chemical potential the ratio of polynomials is preferred,
but we cannot claim this  is the interpolation to use for analytic continuation since systematic differences between finite quark density and finite isospin QCD cannot be excluded. 

\section{The curvatures of the critical lines}
\label{curvatures}

To obtain the curvatures of the critical line at $\mu=0$  ($(d \beta_c(\mu^2) / d \mu^2)|_{\mu=0}$) for isospin and quark chemical potentials respectively,  we tried a common fit to all data  we have collected for the critical couplings at quark chemical potential and isospin chemical potential (in the latter case we included also data at imaginary values):
\begin{equation}
\label{linearfit}
\beta_c(\mu_q,\mu_{\rm iso}) = \beta_c(0) 
+ a_q \left(\frac{\mu_q}{\pi T}\right)^2 
+ a_{\rm iso} \left(\frac{\mu_{\rm iso}}{\pi T}\right)^2 \,.
\end{equation}
We included in the fit as many data as to have a reasonable $\chi^2/{\rm d.o.f.}$ and we obtained: 
$a_q = -0.3997(87)$, $a_{\rm iso} = -0.3606(67)$, $\beta_c(0) =5.32370(57)$ with a $\chi^2/{\rm d.o.f.} = 0.93$.
Therefore we can conclude that the curvatures of the critical lines respectively for isospin and quark chemical potentials differ up to 4 standard deviations.
By expressing the curvatures in terms of dimensionless quantities~\cite{D'Elia:2009tm}
\begin{equation}
\label{RqRiso}
\frac{T_c(\mu_q,\mu_{\rm iso})}{T_c(0)} = 1 
+ R_q \left(\frac{\mu_q}{\pi T}\right)^2 
+ R_{\rm iso} \left(\frac{\mu_{\rm iso}}{\pi T}\right)^2 \,
\end{equation}
with
\begin{equation}
\label{rrr}
R_{q/{\rm iso}} = 
\left. - \frac{1}{a} \frac{\partial\ a}{\partial \beta} \right|_{\beta_c(0)} 
a_{q/{\rm iso}} =
\sqrt{\frac{N_c}{2 \beta_c(0)^3}} 
\frac{1}{\beta_L(\beta_c(0),m_q)} a_{q/{\rm iso}} \,,
\end{equation}
where $\beta_L=a \frac{\partial g_0}{\partial a}$ is the 2-loop lattice beta-function,
we get
$R_q = -0.515(11)$ and $ R_{\rm iso} = -0.465(9)$
in agreement with Refs.~\cite{deForcrand:2003hx,PhysRevD.83.114507,Kogut:2004zg}.
Therefore:
\begin{equation}
\label{ratiocurv}
R_{q - {\rm iso}} = \frac{R_q - R_{\rm iso}}{R_q} = \frac{a_q - a_{\rm iso}}
{a_q} = 0.098(26) \sim 10\% \,.
\end{equation}
This could be the first evidence of the ${\cal{O}}(1/N_c^2)$ difference between the two theories at small chemical 
potential~\cite{Toublan:2005rq,Hanada:2011ju,Armoni:2012jw,Hanada:2012es}.
\begin{figure}[tbp]
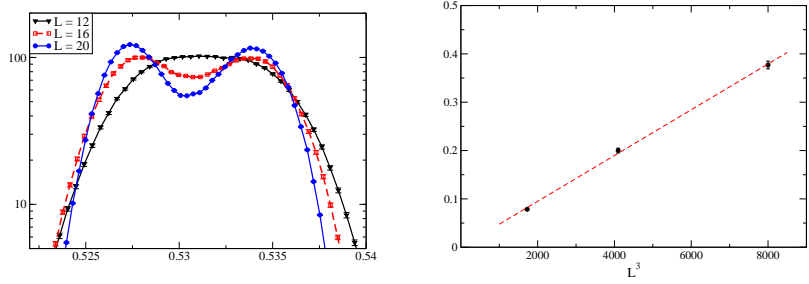
 
\centering
\begin{tabular}{cc}
\includegraphics[width=.35\textwidth,clip]{./FIGURES/histo_0.475.eps} &
\includegraphics[width=.35\textwidth,clip]{./FIGURES/suscmax.eps} 
\end{tabular} 
\caption{(Left) Normalized plaquette distributions 
at the pseudocritical coupling for different
spatial lattice sizes and $\mu_{\rm iso}/(\pi T) = 0.475i$.
(Right) Maxima of the plaquette susceptibility as a function of the 
spatial volume for $\mu_{\rm iso}/(\pi T) = 0.475i$.}
\label{fig5}
\end{figure}

\section{Order of the phase transition at imaginary chemical potentials}
\label{phase}
The phase structure at finite $T$ and imaginary chemical potential may be important of its own and teach us something about the nonperturbative properties of QCD also at zero and small chemical potential. The phase transition at the Roberge-Weiss endpoint could in principle have influence also far from the endpoint.
For $n_f=2$ the Roberge-Weiss transition is first order for small and high quark masses and second order for intermediate quark masses~\cite{Bonati:2010gi}.
In the present study we have $am = 0.05$ so that we expect a second order Roberge-Weiss phase transition (at $\mu_{\text{IM}}/(\pi T)=1/3$) in the case of the quark chemical potential.
On the other hand we expect that imaginary isospin chemical potential may strengthen the transition as an imaginary quark potential does: a first-order transition could be manifest along the pseudocritical line (even for our quark mass value $am=0.05$).
In fig.\ref{fig5} (left) we display the normalized plaquette distributions at the pseudocritical coupling for different spatial lattice sizes ($L_s=12,16,20$), while in fig.\ref{fig5} (right) we can see the 
maxima of the plaquette susceptibility that  scale linearly with the spatial volume. Therefore we conclude that
for $n_f=2$ staggered fermions of mass $am=0.05$ the transition is first order at  $\mu_{\text{iso}}/(\pi T)=0.475i$   and there is possibly a critical point along the line at some smaller value of $\mu_{\text{iso}}/(\pi T)$. Such non-trivial behavior resembles what happens 
for quark chemical potentials~\cite{D'Elia:2009qz,deForcrand:2010he,Bonati:2010gi,Bonati:2012pe}
and may have consequences on the general structure of the QCD phase diagram.

\providecommand{\href}[2]{#2}\begingroup\raggedright\endgroup

\end{document}